\documentclass{emulateapj}

\usepackage{graphicx}

\begin{document}

\newcommand{\Mstar}{\ensuremath{M_{\ast}}} 
\newcommand{\Msun}{\ensuremath{M_{\odot}}}

\newcommand{\Rstar}{\ensuremath{R_{\ast}}}
\newcommand{\Rkep}{\ensuremath{R_{\rm K}}}
\newcommand{\Ralf}{\ensuremath{R_{\rm A}}}
\newcommand{\Resc}{\ensuremath{R_{\rm E}}}
\newcommand{\Rsun}{\ensuremath{R_{\odot}}}

\newcommand{\estar}{\ensuremath{\eta_{\ast}}}
\newcommand{\Beq}{\ensuremath{B_{\rm eq}}}
\newcommand{\Mdot}{\ensuremath{\dot{M}}}

\newcommand{\Vinf}{\ensuremath{V_{\infty}}}
\newcommand{\Veq}{\ensuremath{V_{\rm eq}}}
\newcommand{\Vcrit}{\ensuremath{V_{\rm crit}}}
\newcommand{\vphi}{\ensuremath{v_{\phi}}}

\def\ltwig{\mathrel{\spose{\lower 3pt\hbox{$\mathchar"218$}}
       \raise 2.0pt\hbox{$\mathchar"13C$}}}
\def\gtwig{\mathrel{\spose{\lower 3pt\hbox{$\mathchar"218$}}
       \raise 2.0pt\hbox{$\mathchar"13E$}}}
\def\spose#1{\hbox to 0pt{#1\hss}}

\newcommand{\beq}{\begin{equation}}
\newcommand{\eeq}{\end{equation}}
\newcommand{\beqa}{\begin{eqnarray}}
\newcommand{\eeqa}{\end{eqnarray}}
\def\Exp{{e}} 
\def\=={{\equiv}} 
\def\vinf{v_{\infty}} 
\def\solar{\odot} 
\def\Mdot{\dot M} 
\def\mdot{\dot M} 
\def\zp{{z^{\prime}}} 
\def\rp{{r^{\prime}}}

\title{The Effect of Porosity on X-ray Emission Line Profiles from
Hot-Star Winds}  
\shorttitle{Effect of Porosity on X-ray Line Profiles} 

\email{owocki@bartol.udel.edu\\ cohen@astro.swarthmore.edu}

\author{Stanley P. Owocki}
\affil{Bartol Research Institute, Department of Physics \& Astronomy,
        University of Delaware, Newark, DE 19716}
\author{David H. Cohen}
\affil{Department of Physics \& Astronomy, Swarthmore College, Swarthmore, PA 19081}

\begin{abstract}
We investigate the degree to which the nearly symmetric form of X-ray 
emission lines seen in {\it Chandra} spectra of
early-type supergiant stars could be explained by a possibly
porous nature of their spatially structured stellar winds.
Such porosity could effectively reduce the bound-free absorption of 
X-rays emitted by embedded wind shocks, 
and thus
allow a more similar transmission of red- vs.\ 
blue-shifted emission from the back vs.\ front hemispheres.
To obtain the localized self-shielding that is central to this porosity
effect, it is necessary that the individual clumps be optically
thick.
In a medium consisting of clumps of size $\ell$ and volume filling
factor $f$, we argue that the general modification in effective
opacity should scale 
approximately as $\kappa_{eff} \approx \kappa/(1+\tau_{c})$,
where, for a given atomic opacity $\kappa$ 
and mean density $\rho$, the clump optical thickness scales as 
$\tau_{c} = \kappa \rho \ell/f$.
For a simple wind structure parameterization in which the `porosity length'
$h \equiv \ell/f$ 
increases with local radius $r$ as $h = h' \, r$, 
we find that a substantial reduction in wind absorption
requires a quite large porosity scale factor, $h' \gtwig 1$, 
implying large porosity lengths $ h \gtwig r$.
The associated wind structure must thus have either a relatively
large scale $ \ell \ltwig r$, or a small volume filling factor
$f \approx \ell/r \ll 1$, or some combination of these.
We argue that 
the relatively small-scale, moderate compressions generated by intrinsic
instabilities in line-driving
are unlikely to give such large porosity lengths.
This raises questions about whether porosity effects could play a 
significant role in explaining nearly symmetric X-ray line profiles,
leaving again the prospect of instead having to invoke a substantial 
(ca.\ factor 5) downward revision in the assumed mass-loss rates.
\end{abstract}


\section{Introduction} \label{sec:intro}

The high sensitivity and high spectral resolution of spectrometers on 
the {\it Chandra} X-ray observatory have made it possible to resolve X-ray
emission line profiles from several hot, bright supergiant stars, e.g.
$\zeta$~Pup,
$\zeta$~Ori,
$\delta$~Ori, 
$\epsilon$~Ori,
$\iota$~Ori,
$\xi$~Per,
and
Cyg~OB2\ 8A.
[For an overview,
see the introduction and discussion sections in
\citet{Cohen2006}.]
The general broadness of these emission lines, with velocity
half-widths of ca.\ 1000~km~s$^{-1}$, is generally consistent with the idea 
that the X-rays are emitted in the expanding, highly supersonic
stellar wind, perhaps from embedded shocks generated by instabilities 
associated with the line-driving of the overall wind outflow.
However, these profiles are also generally quite symmetric\footnote{
Actually, even for stars that have been qualitatively characterized as having
symmetric X-ray line profiles,
quantitative analyses \citep[e.g.,][]{Cohen2006} 
show there is both a distinct blue-shift in the line centroid and 
a net skewness in the profile shape, 
albeit at levels much less than expected from standard wind models.}
between the red and blue side, implying a small degree of attenuation 
of the red-side emission thought to originate in the back hemisphere
relative to the observer.

In the standard wind-shock picture, the X-ray emission is believed  to
come from only a small fraction ($<$ 1\%) of the gas 
\citep{OCR88, F95, Fetal97},
with the bulk of the wind consisting of relatively cool material 
with a substantial X-ray absorption opacity from bound-free transitions of
helium 
and heavier ions.
With the standard mass-loss rate for these stars, 
which are derived from either H$\alpha$ or free-free radio emission, 
the  characteristic bound-free optical depths along a radial ray to the 
surface are expected to be of order ten or more
\citep{Hillier1993};
since this implies a substantial attenuation of red-shifted emission 
originating from the back hemisphere, the expected X-ray emission line
profiles have a markedly asymmetric form, with a much stronger blue side
and a lower, more attenuated red side
\citep[][hereafter OC]{OC01}.
Within a simple parameterized model, fitting the more symmetric
observed profiles has thus required a substantial 
(ca.\ a factor 5 or more!) 
reduction in the assumed wind mass-loss rates 
\citep{KCO03,Cohen2006}.
If confirmed, such a radical reduction in supergiant mass loss would have
far-reaching consequences for both massive star evolution and the
broad influence of wind mass loss on the structure of the interstellar
medium.

This paper investigates an alternative scenario in which the reduction
in wind attenuation might instead result from a spatially {\em porous}
nature of the stellar wind.
If wind material is compressed into localized, optically thick clumps, 
then red-shifted emission from the back hemisphere might be more
readily transmitted through the relatively low-density
channels or porous regions between the clumps.

\citet{FOH03} and \citet{OFH04} have in fact examined such effects
in quite detailed models that assume a specific `pancake' form for 
the dense structures,
under the presumption that these would arise naturally from the strong radial
compressions associated with the intrinsic instability of the
line-driving of such hot-star winds.
Although 1D simulations of the nonlinear evolution of the instability
do lead to compression into geometrically thin shells  \citep{OCR88, F95}, 
recent 2D models 
\citep{DO03, DO05} suggest the structure may instead break into clumps with 
a similarly small lateral and radial scale.
But even such initial 2D simulations do not yet properly treat the
lateral radiation transport that might couple material, and so the scale,
compression level, and degree of anisotropy of instability-generated
structure in a fully consistent 3D model is still uncertain.

This paper thus develops a simpler parameterized approach that aims to
identify the basic properties needed for porosity to have a
significant effect on the overall attenuation, and particularly 
to allow the near-symmetry of observed X-ray emission profiles.
It builds on the simple parameterization developed by OC,
which has been successfully applied to
derive key wind properties needed to fit observed X-ray profiles under the 
assumption that absorption follows the usual (non-porous) form for a
smooth wind
\citep{KCO03, Cohen2006}.

Based on recent analyses of porosity effects in moderating
continuum-driven mass loss \citep{OGS04},
we present here (\S 2) simple scalings for the porosity reduction of 
effective opacity as a function of the optical thickness $\tau_{c}$ of
individual clumps in a structured flow;
for a given mean density and (microscopic) opacity, we show that this 
depends on the ratio of the clump scale $\ell$ to volume filling
factor $f$, a quantity we dub the `porosity length',
$h \equiv \ell/f$.
We next apply (\S 3) this scaling for porosity-modified opacity
within the spatial integration for the wind optical depth,
showing that, 
for a physically reasonable model in which this porosity length is assumed 
to increase linearly with local radius as $h = h' r$,
the integral can evaluated analytically 
(assuming also a canonical `beta=1' form for the wind velocity law).
We then use (\S 4) this analytic optical depth to compute X-ray emission line
profiles for the simple case that the wind X-ray emission has a constant
filling factor above some initial onset radius, set here to 
$R_{o} = 1.5 \Rstar$, roughly where instability simulations show the initial
appearance of X-ray emitting shocks, and roughly consistent with
values derived by fits to the X-ray data.
The discussion section (\S 5) then reviews implications of the key
result that obtaining nearly symmetric profiles from an otherwise
optically thick wind requires very large porosity lengths, $h \gtwig r$,
a requirement that seems at odds with the small scale and moderate
compression factors found in instability simulations.
The conclusion (\S 6) summarizes these results and briefly discusses
the potential for future application of our porosity parametrization
in spectroscopic analysis tools.

\vspace{1.2cm}
\section{Clumping vs.\ Porosity Effects in a Structured Medium} 
\label{sec:PorVsClump}

\subsection{Density-Squared Clumping Correction}  \label{subsec:Densq}

Before discussing how porosity effects can alter 
diagnostics like bound-free absorption
that scale linearly with density,
it is helpful first to review briefly
the usual account of how the clumping of a medium can 
alter diagnostics that scale with the {\em square} of the density.
For example, emission and absorption from atomic states that arise from
recombination, collisional excitation, or free-free processes 
all depend on the proximate interaction of two constituents, 
e.g.\ electrons and ions, and thus scale with the product of 
their individual particle density, e.g.\ $n_{e} n_{i}$, which for a
fixed ionization and abundance is simply proportional to 
the square of the mass density, $\rho^{2}$.
The effect of spatial structure on such diagnostics is thus
traditionally accounted for in terms of a simple density-squared
clumping correction factor,
\beq
C_{c} \equiv { \left < \rho^{2} \right > \over \left < \rho \right >^{2} }
\, ,
\label{fcdef}
\eeq
where the angle brackets denote a volume averaging.
For example, in a simple model in which 
the medium consists entirely of
clumps of scale $\ell$ and mass $m_{c}$ 
that are separated by a mean distance $L \gg \ell$, 
the mean density is $\left <\rho \right > = m_{c}/L^{3}$, 
whereas the individual clump density is
$\rho_{c} = m_{c}/\ell^{3} = \left < \rho \right > (L/\ell)^{3}$.
Application in eqn.\ (\ref{fcdef}) then implies that the clumping
correction is just given by the inverse of the volume filling factor,
\beq
C_{c} =  { L^{3} \over \ell^{3} } \equiv { 1 \over f }
\, .
\label{fvdef}
\eeq
For diagnostics of wind mass-loss rate, e.g.\ Balmer or radio
emission, the associated overestimate in inferred mass loss  
scales as ${\dot M} \sim \sqrt{C_{c}} \sim 1/\sqrt{f}$.

A key point here is 
that this density-squared clumping correction depends 
only on the volume filling factor, $f = \ell^{3}/L^{3}$,
and {\em not} on the scale $\ell$ of individual clumps.
As long as the emission can escape from each local emitting clump
(i.e., the clumps remain optically thin),
the correction factor thus applies to structure ranging,
for example, 
from very small-scale instability-generated turbulence
\citep{DO03, DO05},
to possible stellar-scale magnetically confined loops
\citep{UO02}.

\subsection{Porosity Reduction in Linear-Density Opacity for Optically
Thick Clumps} 
\label{subsec:PorThick}

The attenuation of X-rays emitted within a stellar wind occurs through 
bound-free absorption, primarily from the ground-state.
Since this is the dominant stage of the absorbing ions, and exists
independent of interaction with other particles, 
the associated absorption scales only {\em  linearly} with 
density,
with the volume opacity (attenuation per unit length) 
given by
$\chi = \kappa \rho$,
where the mass opacity $\kappa$ (mass absorption coefficient) has
units of a cross section per unit mass, e.g.\ $cm^{2}/g$ in CGS.
Such linear-density absorption is often thought of as being
unaffected by clumping.

If, however, we consider the above clump model in the case where the
individual clumps are {\em optically thick}, then the 
`effective opacity' of the clump ensemble
can be written in terms of the ratio of the physical cross
section of the clumps to their mass,
\beq
\kappa_{eff} \equiv { \ell^{2} \over m_{c} }
= { \kappa \over \tau_{c}}
~~~ ; ~~~ \tau_{c} \gg 1
\, .
\label{keffthick}
\eeq
The latter equality shows that, 
relative to the atomic opacity $\kappa$, 
this effective opacity is reduced by a factor that scales with the 
inverse of the clump optical thickness,
$\tau_{c} = \kappa \rho_{c} l = \kappa \left < \rho \right >l/f$.

The above scaling  serves to emphasize a key requirement for
porosity, namely the {\em local self-shielding} of material within optically
thick clumps, allowing then for a more transparent transmission of
radiation through the porous interclump channels.

Note then 
that the clump optical thickness that determines 
the effective opacity reduction depends on the {\em ratio} of the clump 
{\em scale} to the volume filling factor, a quantity which we call 
the porosity length, $h \equiv \ell/f$.
This represents an essential distinction between the porosity effect
and the usual density-squared clumping correction, which as noted
above depends only on the volume filling factor without 
any dependence on the clump size scale.

\section{General Porosity Law Bridging Optically Thin and Thick Clump
Limits}

To generalize the above effective opacity to a scaling that applies to
both the optically thick and thin limits, consider that the effective
absorption of clumps is more generally set by the geometric cross section 
times a correction for the net absorption fraction, 
$\sigma_{eff} = \ell^{2} [1- \exp(-\tau_{c}) ]$.
Applying this to modify the scaling in eqn.\ (\ref{keffthick}),
we obtain a general porosity reduction in opacity of the form
\citep{OGS04},
\beq
{ \kappa_{eff} \over \kappa } = { 1 - e^{-\tau_{c}} \over \tau_{c} }
\, .
\label{keffabs}
\eeq
In the optically thick clump limit  $\tau_{c} \gg 1$, this gives the reduced opacity  
$\kappa_{eff}/\kappa \approx 1/\tau_{c}$
of eqn.\ (\ref{keffthick}),
while in the optically thin  limit  $\tau_{c} \ll 1$, it
recovers the atomic opacity $\kappa_{eff} \approx \kappa$.

\begin{figure}
\begin{center}
\epsscale{1.15}
\plotone{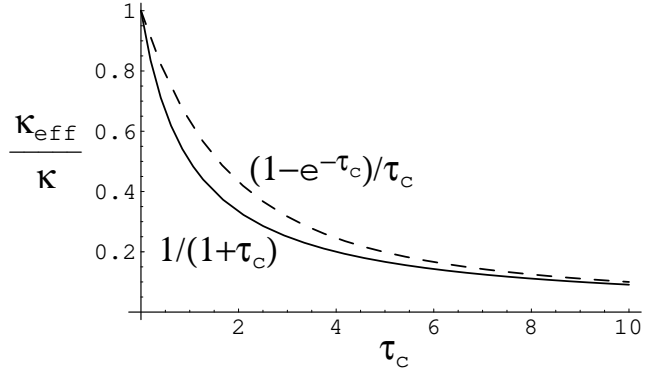}
\caption{
Comparison of absorption [dashed upper curve; eqn.\ (\ref{keffabs})] 
and mean-path [solid lower curve; eqn.\ (\ref{keffmfp})]
scalings of effective opacity in a porous medium,
plotted as a function of clump optical thickness $\tau_{c}$.
} 
\label{fig1}
\end{center}
\end{figure}

An even simpler, alternative bridging form can be derived by focussing on the
effective mean path length within the medium, which scales with the
inverse of the effective volume opacity, 
$1/\kappa_{eff} \left < \rho \right >$.
Within a model in which such an effective opacity adds in the inverse
of contributing components (much as Rosseland mean opacity defined for
weighting frequency-averaged opacity),
we add the microscopic and clump components of path length as,
\beq
{1 \over \kappa_{eff} \left < \rho \right > } =
{1 \over \kappa \left < \rho \right > }  + h 
\, ,
\label{pathadd}
\eeq
where we note that the porosity length defined above also defines a
mean free path between clumps,
$h \equiv \ell/f = L^{3}/\ell^{2}$.
This scaling solves to a general effective opacity of the form,
\beq
{ \kappa_{eff} \over \kappa } = { 1  \over 1 +\tau_{c} }
\, .
\label{keffmfp}
\eeq
This again gives both the correct scalings in the opposite 
asymptotic limits of optically thin vs.\ thick clumps.
For moderately small clump optical depths $\tau \ltwig 1$, 
Taylor expansion shows the reduction is somewhat steeper for 
this new mean-path form of eqn.\ (\ref{keffmfp}), i.e.
as $1-\tau_{c}$ instead of the slightly 
weaker $1-\tau_{c}/2$ for the absorption scaling of
eqn.\ (\ref{keffabs}).
But the plot in Fig. \ref{fig1} shows that both forms have a very similar 
overall variation with clump optical depth.

\section{Porosity Effect on Wind Optical Depth}
\label{sec:poroptdep}

The above formalism provides a convenient way to explore the effect of
porosity on wind attenuation of X-ray emission.
Our basic approach here is to generalize the parameterized
analysis of OC
to include the porosity reduction in
effective absorption. 
For the effective opacity, we choose to work with the slightly simpler 
mean-path form (\ref{keffmfp}),
which avoids the complicating effects of the exponential function, 
with the clump optical thickness correction only appearing in a single 
linear term in the denominator.
Following eqn.\ (OC-2),
the 
effective 
optical depth to a position $z$ along a ray with impact parameter $p$
is now written as,
\begin{equation}
t_{eff} [p,z] 
= \int_z^\infty 
{\kappa \rho [ r' ] \over 1 + \kappa \rho[r'] h[r']} \, dz' 
\, ,
\label{taudef}
\end{equation}
where $\rho [\rp] $ is the smoothed-out mass density at radius 
$r' \equiv \sqrt{p^2+z'^2}$,
and $h[r']$ is the (possibly radially dependent) porosity length.

For a steady-state wind with a simple (`beta=1') velocity law of the
form $w(r) \equiv v(r)/\vinf = (1-\Rstar/r) \equiv y(r)$, 
eqn.\ (\ref{taudef}) becomes [cf.\ eqn.\ (OC-4)],
\begin{equation}
t_{eff} [p,z] =
\tau_\ast
\int_z^\infty { \Rstar dz' \over r' (r'-\Rstar) + \tau_{\ast} h[r']}
 \, ,
\label{taupz}
\end{equation}
where, as in OC,
we have here used the mass-loss rate, $\Mdot \equiv 4 \pi \rho v r^2$,
to define a characteristic wind optical 
depth\footnote{Note
that in a smooth wind with a constant velocity
$v=\vinf$, the radial ($p=0$) optical depth at radius $r$ would be
given simply by $t[0,r] = \tau_\ast \Rstar/r$.
Thus in such a constant-velocity wind, $\tau_\ast$ would be the radial
optical depth at the surface radius $\Rstar$, while 
$R_1 = \tau_\ast \Rstar$ would be the radius of unit radial optical
depth.},
$\tau_\ast \equiv \kappa \Mdot /4 \pi \vinf \Rstar$.
As in eqn.\ (OC-4), for rays intersecting the core ($p \le \Rstar$),
eqn.\ (\ref{taupz}) is restricted to locations in front of the star, 
i.e.\ $z > \sqrt{\Rstar^{2} - p^{2}}$, since otherwise the optical
depth becomes infinite, due to absorption by the star.

To account for the likely expansion of the clump size with the overall
wind expansion,
let us specifically assume here 
that the porosity length increases linearly with the radius, i.e.\ as $h = h' r$.
Fortunately, for this physically quite reasonable case, 
the integral in eqn.\ (\ref{taupz}) can be evaluated analytically to
give [cf.\ eqn.\ (OC-5)]
\begin{equation}
{t_{eff}[p,z] \over \tau_\ast} = 
{ 
\left [ 
\arctan \left ( { (1-\tau_{\ast} h') z' \over r' z_{h}/\Rstar } \right) 
+
\arctan \left ( { z' \over z_{h} } \right ) 
\right ]^{z'\rightarrow \infty}_{z'=z} 
\over z_{h}/\Rstar } 
\, ,
\label{taupzb1}
\end{equation}
where $z_{h} \equiv \sqrt{p^{2} - \Rstar^{2} (1-\tau_{\ast} h')^{2}}$.
In terms of the local direction cosine $\mu = z/r$, 
the effective optical depth as a function of spherical coordinates ($\mu,r$) 
is given by
\begin{equation}
\tau_{eff}[\mu,r] = t_{eff} \left[ \sqrt{1-\mu^2} \, r, \, \mu r \right]
\, .
\label{t2tau}
\end{equation}

At any given radius the projected Doppler shift from the wind
velocity depends only on direction cosine $\mu$.
Thus, if we assume the intrinsic line profile of the local wind emission 
has a narrow, delta-function form, then
upon integration over emission direction,
we find a simple transformation from direction cosine
to Doppler-shifted wavelength,
$\mu \rightarrow -x/w(r)$,
where 
$w(r) = v(r)/\vinf = 1 - \Rstar/r$ is the scaled velocity law,
and 
$ x \equiv (\lambda/\lambda_o -1) c/\vinf$
is the Doppler shift in wavelength $\lambda$ from line center 
$\lambda_{o}$, measured in units of the wind terminal speed $\vinf$.

\begin{figure}
\begin{center}
\epsscale{1.15}
\plotone{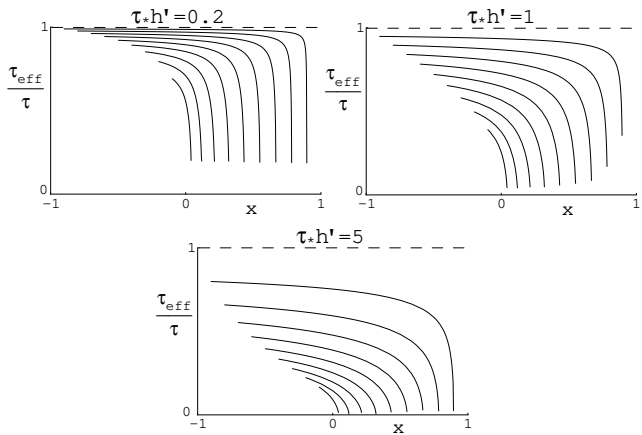}
\caption{
Ratio of the porosity-reduced effective optical depth to 
optical depth in a smooth wind,
$\tau_{eff}[-x/w(r),r]/\tau [-x/w(r),r]$, 
plotted vs.\ the 
scaled wavelength $x$, at selected radii $r$  
in porosity models with 
$\tau_{\ast} h'=$ 0.2, 1, and 5 (left top to bottom panels).
For each panel, 
the nine overplotted curves represent equal increments of 
scaled inverse-radius coordinate from $y \equiv 1-\Rstar/r=$~0.1 to 0.9, 
ordered from lower left to upper right.
}
\label{fig2}
\end{center}
\end{figure}

Figure \ref{fig2} plots the porosity reduction in optical depth vs.\ 
wavelength $x$, for a selection of source radii $r$,
with the  panels from left top to bottom assuming increasing porosity,
parameterized by $\tau_{\ast} h' = $ 0.2, 1, and 5.
In the top-left panel,  steep reductions only occur near those red-side
wavelengths that, for the given source radius, require ray passage 
through the very inner wind, where the high density makes the clumps optically
thick even for the modest assumed porosity length;
since this  generally means the overall optical depth in these regions
is also quite high, the reductions still do not make the regions
very transparent.
But for the increasing porosity cases in the middle and lower panels, the
reduction in optical depth occurs over a much broader range of
wavelengths;
these large porosity cases can thus indeed lead to a  more transparent wind,
with, as we now show, notable changes in the line profile.

\vspace{0.5cm}
\section{Porosity Effect on X-ray Emission Line Profiles}
\label{sec:porxprof}

With the optical depths in hand, the computation of the emission line 
profiles follows directly the approach by OC.
The wavelength-dependent X-ray luminosity can then be
evaluated by straightforward numeral integration 
of a single integral in the scaled inverse-radius coordinate 
$y \equiv 1-\Rstar/r$
[cf.\ eqn.\ (OC-9)],
\begin{equation}
L_x 
\, \propto \,  
\int_{y_x}^1 \,  
{ dy
\over y^{3}}
\, \Exp^{-\tau_{eff} \left[ -x/y, \Rstar/(1-y) \right]} \, ,
\label{lxq}
\end{equation}
where $y_x \equiv  \max[|x|,1-\Rstar/R_{o}]$.
For simplicity, we have assumed here that the X-ray emission filling 
factor is zero below a minimum X-ray emission radius $R_{o}$, 
and constant above this (i.e.\ the OC $q=0$ case).
Specifically, we assume here an X-ray emission onset radius of 
$R_{o} = 1.5 \Rstar$, roughly where instability simulations show the
appearance of self-excited wind structure with embedded shocks
\citep[e.g.,][]{RO02}, 
and roughly the favored value in detailed parameter fits to observed X-ray
spectra 
\citep{KCO03, Cohen2006}.

\begin{figure}
\begin{center}
\epsscale{1.15}
\plotone{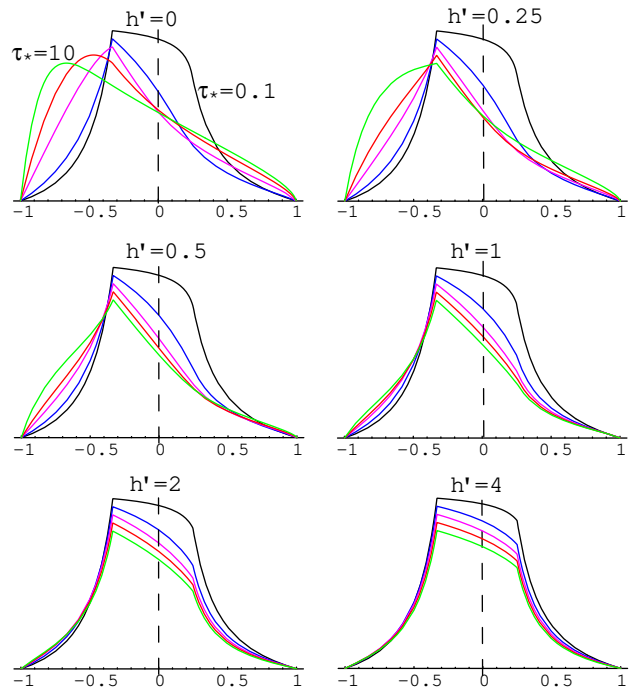}
\caption{
X-ray line profiles vs.\ scaled wavelength 
$x \equiv (\lambda/\lambda_o -1) c/\vinf $, 
overplotted in each panel for optical depth parameters 
$\tau_{\ast} =$ 0.1, 1, 3, 5, and 10 
(black, blue, violet, red, green), and
normalized to have peaks decrease by 5\% for each step in
$\tau_{\ast}$.
The panels compare results for various porosity scale factors
$h'=$ 0, 0.25, 0.5, 1, 2, and 4, 
ordered from upper left to lower right.
The vertical dashed line marks the line center. 
Note that porosity can make otherwise optically thick cases (i.e.
$\tau_{\ast} = 3, 5, 10$) have nearly symmetric profiles, but only
with quite large porosity scale factors, $h' > 1$, as seen in the
lowermost panels.
}
\label{fig3}
\end{center}
\end{figure}

Fig. \ref{fig3} compares line profiles for various porosity scale
factors, from the no porosity case ($h'=0$; upper left)
to large factors ($h'=4$; lower right).
Note that the profiles with large optical depth $\tau_{\ast} = 10$
(green curves, scaled to be the lowest in each panel) 
have strong blue-shifted asymmetry for cases with no or moderate porosity,
($h' \le 1$), and approach the near symmetry of the optically thin case 
($\tau_{\ast} = 0.1$; black curves, scaled to have the highest peaks)
only for models with very large porosity scale factor, $h' = $2 or 4
(lowermost panels).

\section{Discussion}
\label{sec:disc}

The basic result here is thus that, 
for cases with a large overall wind optical depth ($\tau_{\ast} > 2$),
achieving a near symmetry in emission line profiles requires very 
large porosity lengths, $h \gtwig r$. 
Since $h \equiv \ell/f$, this implies that the wind structure must
either have a very large scale, $ \ell \ltwig r$, or a small filling
factor, $f < \ell/r$, or some combination of these.

It is interesting to compare this result with those obtained
by \citet{FOH03} and \citet{OFH04} for their more specialized 
fractured-wind models that assume the line-driven instability will lead to
radially compressed pancake structures separated by strong rarefactions.
These authors 
have generally argued that the porous regions between
such radially compressed structures 
could allow transmission of emission from the back hemisphere, 
and thus explain the greater-than-expected symmetry of observed X-ray 
line profiles.
However, their most recent efforts
\citep{OFH05}
to obtain symmetric profiles with this model have
typically assumed quite large radial separations between the
dense compressions, on the order of a stellar radius or more.

This seems generally consistent with the requirement here for large
porosity lengths.
Indeed, in this picture of pancakes that arise from 1-D, radial
compressions, the volume filling factor is just given by $f = \ell/L$,
where $\ell$ is now the {\it radial} compression size of the pancakes,
and $L$ is the radial separation between them. 
Then using our definition for the porosity length, we see
that in this case this porosity length is indeed just given by the 
separation scale, $h = \ell/f = L$.
Moreover, if the pancakes are optically thick, this also gives a typical
mean-free-path through the porous wind, much as in our description
above.
In both analyses, we see then that, for porosity to be sufficient
to make the wind transparent, this mean path has to be quite large, 
comparable to or larger than a stellar radius.

Overall, these results thus raise serious issues for associating significant 
porosity reduction in X-ray absorption with the small-scale wind structure 
expected from the intrinsic instability of line driving.
In linear analyses
\citep[e.g.,][]{OR84},
this instability occurs for radial perturbations on the
scale of the Sobolev length, 
$L_{Sob} = v_{th}/(dv/dr) \approx \Rstar v_{th}/\vinf $;
since typical wind terminal speeds of order
$\vinf \approx 1000$~km/s 
are much larger than the typical ion thermal speed 
$v_{th} \approx 10$~km/s,
we find $L_{Sob} \approx 0.01 \Rstar$.
In 1D nonlinear simulations, this is indeed the typical scale of
initial structure in the inner wind, with some subsequent outward increase 
due to merging as faster shells collide with slower ones
\citep[e.g.,][]{F95, RO02}.
But in 2D models \citep{DO03, DO05}, individual clumps with different 
radial speeds can pass by each other, with shearing effects competing 
with merging, so that the typical scale remains quite small.
In both 1D and 2D simulations,
the net compression is found to be quite moderate, with associated volume
filling factors typically of order $f \approx 0.1$.
Thus to the extent that the complex structure can be characterized by 
a single porosity length, it seems this would be of order 
$h = 0.01 \Rstar/0.1 \approx 0.1 \Rstar$, which is much smaller than 
what the above analysis suggests  is necessary to give a substantial porosity 
effect on line profiles.

Of course, compared to the complex structure in such instability
simulations,
the above clump model is highly idealized and even simplistic,
characterizing the structure in terms of a single size scale, and
essentially assuming that the porous regions in between the isolated
clumps are completely empty and thus transparent.
But such simplifications would generally seem only to favor the development of
porous transport, representing a kind of best-case scenario;
it thus seems quite significant that even in this case the requirements 
for achieving a significant porosity are actually quite stringent.

The more extensive porosity analysis by \citet{OGS04} suggests, for example,
that in a medium with a distribution of clump scales, the porosity
reduction in opacity scales with a weaker-than-linear power of inverse
density, e.g.\ $1/\rho^{\alpha}$ for power-law distribution with
positive index $\alpha < 1$, which
reflects self-shielding as different scale clumps become optically thick.
In their `inside-out' context of porosity-mediated mass loss from
dense layers of the stellar envelope, this modification
tends to shift the wind sonic point to higher density, thus serving to
increase the derived mass loss.
But in the present `outside-in' context of wind attenuation of
X-rays seen by an external observer, the weaker scaling will only
make it more difficult to give the outer regions near the X-ray
photosphere a significant porosity reduction in absorption.

Thus, despite the simplicity of our basic model and analysis, we
believe the central physical arguments are quite general and robust, 
indicating that significant porosity reductions in the absorption of otherwise
optically thick winds are only possible for large porosity lengths
$h \equiv \ell/f$, 
implying either quite large size scales $\ell$,
or strong compressions into a small filling factor $f$, 
or some combination of these.
While large-scale structures may indeed exist in 
the winds of some specific stars, 
for example owing to global magnetic fields, 
the kind of ubiquitous structure expected from the 
intrinsic instability of line-driving 
seems simply to have too small a scale to play much role in porosity
reduction of wind absorption.

Moreover, sufficiently large structure would seem likely to be associated
with temporal variability of emergent X-rays due to, e.g., 
rotational modulation of the attenuation 
(in addition to likely modulations or even intrinsic variations in 
emission);
but such variability is not generally seen in the X-ray spectra 
of hot stars, 
which are typically found to be constant to a level of a few percent or so
Ê\citep{Berg96, CCM97}.
In addition, because magnetic fields generally retard or confine the wind
outflow, leading to high-density structures 
\citep{UO02}, their overall effect could 
even be an overall {\em increase} in absorption, 
relative to what would occur in a smooth, spherically symmetric wind outflow.
Taken together, these considerations seem to argue against an important
porosity effect from wind structure of any scale.

Of course, as extensively discussed in the literature,
even small-scale, optically thin structure can have a major impact on
diagnostics that scale with density-squared, leading for example 
to overestimates of the wind mass loss that scale with $1/\sqrt{f}$.
If mass-loss rates are thereby revised downward by significant factors
of five or more, then it becomes possible to explain the observed 
near-symmetry of observed X-ray line-profiles 
in terms of reduced overall optical
depths $\tau_{\ast}$
\citep{KCO03, Cohen2006},
without needing to invoke any porosity reduction in the effective
opacity.

Mass loss reductions of this magnitude have in fact been recently
suggested, 
based on NLTE models of hot star-spectra that include transitions
with a mixed contribution from single-density and density-squared 
processes
\citep{LBH05},
and
based on FUSE observations of wind lines from a fuller range of ions 
spanning the dominant stage
\citep{FMP06}.
Such substantial reductions in hot-star mass loss would have
broad-ranging implications, for example for  massive-star
evolution, and for the physics of the interstellar medium.

\section{Summary} \label{sec:summary}

We have carried out a simplified, parameterized analysis of the 
potential role of a porous medium in reducing the effective absorption 
of X-rays emitted in an expanding stellar wind.
In contrast to the usual corrections for density-squared processes,
which depends only on the volume filling factor,
we show that the importance of such porosity effects depend
largely on a quantity we call the porosity length, 
which is set by the ratio of the characteristic clump size scale to
this filling  factor.
This determines the density and opacity for which the individual
clumps become optically thick, leading to a local self-shielding 
within clumps that reduces  the effective opacity of the medium
by a factor that scales with the inverse density.
A key result is that porosity reduction of absorption at a level to
make an otherwise optically thick line emission nearly symmetric
requires very large porosity lengths, on order of the local radius or 
more, $h \gtwig r$.
Because a large porosity length requires a combination of large-size clump
structures or a small, compressed filling factor, 
it seems unlikely that this could result from the
small-scale, moderately compressed structure expected from the
intrinsic instability due to line-driving.
Despite the simplicity of the basic model, the key physical reasons for
requiring large porosity lengths seem quite general and robust.
An overall conclusion is thus that explaining the unexpectedly symmetric 
form for observed X-ray line-profiles may instead require substantial
reductions in inferred mass-loss rates.

The parameterization developed here for the porosity reduction of
opacity (see eqn.\ [\ref{keffmfp})] is simple enough to lend itself to
general application within spectroscopic analysis codes like 
{\it XSPEC}.
Future work on fitting observed spectra could then derive more 
specific requirements for porosity models to match line profiles,
allowing one to explore further the trade-off between invoking porosity
or lowering the wind mass-loss rate.
 
\acknowledgements{DHC acknowledges NASA contract AR5-6003X to 
Swarthmore College through  the {\it Chandra} X-ray Center. 
SPO acknowledges NSF grants AST-0097983 and AST-0507581, and 
NASA/{\it Chandra} Theory grant GO3-3024C.}

\bibliographystyle{aastex}

\begin{thebibliography}
    
\bibitem[Berghoefer et al.(1996)]{Berg96} 
Berghoefer, T.~W., Baade, D., Schmitt, J.~H.~M.~M., Kudritzki, R.-P., 
Puls, J., Hillier,  D.~J., \& Pauldrach, A.~W.~A.\ 1996, \aap, 306, 899 

\bibitem[Bouret et al.(2005)]{LBH05} 
Bouret, J.-C., Lanz, T.,  \& Hillier, D.~J.\ 2005, \aap, 438, 301 

\bibitem[Cohen et al.(1997)]{CCM97} 
Cohen, D.~H., Cassinelli, J.~P., \& Macfarlane, J.~J.\ 1997, \apj, 487, 867

\bibitem[Cohen et al.(2006)]{Cohen2006} 
Cohen, D.~H., Leutenegger, M.~A., Grizzard, K.~T., Reed, C.~L., Kramer,
R.~H., and Owocki, S.~P.\ 2006, \mnras, submitted

\bibitem[Dessart \& Owocki(2003)]{DO03} 
Dessart, L. \& Owocki, S. P.  2003, \aap, 406, L1

\bibitem[Dessart \& Owocki(2005)]{DO05} 
Dessart, L., \& Owocki, S.~P.\ 2005, \aap, 437, 657 

\bibitem[Feldmeier(1995)]{F95} 
Feldmeier, A.\ 1995, \aap,   299, 523 
   
\bibitem[Feldmeier et al.(1997)]{Fetal97} 
Feldmeier, A., Kudritzki, R.-P., Palsa, R., Pauldrach, A. W. A., \& 
Puls, J. 1997, \aap, 320, 899

\bibitem[Feldmeier, Oskinova, \& Hamann(2003)]{FOH03} 
Feldmeier, A.,  Oskinova, L., \& Hamann, W.-R. 2003, \aap, 403, 217

\bibitem[Fullerton, Massa, \& Prinja(2006)]{FMP06} 
Fullerton, A. W., Massa, D. L., \& Prinja, R. K. 2006, 
\apj, 637, 1025
 
\bibitem[Hillier et al.(1993)]{Hillier1993} Hillier, D.~J., Kudritzki,
  R.~P., Pauldrach, A.~W., Baade, D., Cassinelli, J.~P., Puls, J., \&
  Schmitt, J.~H.~M.~M. 1993, \apj, 276, 117
  
\bibitem[Kramer, Cohen, \& Owocki(2003)]{KCO03} 
Kramer, R.~H., Cohen, D.~H., \& Owocki, S.~P.\ 2003, \apj, 592, 532 

\bibitem[Oskinova, Feldmeier, \& Hamann(2004)]{OFH04} 
  Oskinova, L., Feldmeier, A., \& Hamann, W.-R. 2004, \aap, 422, 675

\bibitem[Oskinova, Feldmeier, \& Hamann(2005)]{OFH05} 
Oskinova, L., Feldmeier, A., \& Hamann, W.-R. 2005, 
astro-ph/0511019

\bibitem[Owocki, Castor, \& Rybicki(1988)]{OCR88} 
Owocki, S. P., Castor, J. I., \& Rybicki, G. B.  1988, \apj, 335, 914
    
\bibitem[Owocki \& Cohen(2001)]{OC01} 
Owocki, S. P. \& Cohen, D. H. 2001, \apj, 559, 1108

\bibitem[Owocki, Gayley, \& Shaviv(2004)]{OGS04} 
Owocki, S. P., Gayley, K. G. \& Shaviv, N. J.  2004, \apj, 616, 525

\bibitem[Owocki \& Rybicki(1984)]{OR84} 
Owocki, S. P. \& Rybicki, G. B.  1984, \apj, 284, 337

\bibitem[Runacres \& Owocki(2002)]{RO02} 
Runacres, M. C. \& Owocki,  S. P. 2002, \aap, 381, 1015

\bibitem[ud-Doula \& Owocki(2002)]{UO02} 
ud-Doula, A.,\& Owocki, S. P. 2002, \apj, 576, 413

\end{thebibliography}

\end{document}